\documentclass[12pt]{iopart}

\usepackage{iopams}  
\usepackage{graphicx}
\begin{document}

\title{Topological Objects in Holographic QCD}

\author{Hideo Suganuma}

\address{Department of Physics, Kyoto University, \\
Kitashirakawaoiwake, Sakyo, Kyoto 606-8502, Japan}
\ead{suganuma@scphys.kyoto-u.ac.jp}

\author{Keiichiro Hori}

\address{Department of Physics, Kyoto University, \\
Kitashirakawaoiwake, Sakyo, Kyoto 606-8502, Japan}
\ead{hori@ruby.scphys.kyoto-u.ac.jp}

\vspace{10pt}
\begin{indented}
\item[]March 2020
\end{indented}

\begin{abstract}

We study topological objects in holographic QCD based on the Sakai-Sugimoto model, which is constructed with $N_c$ D4 branes and $N_f$ D8/$\bar{\rm D8}$ branes in the superstring theory, and is infrared equivalent to 1+3 dimensional massless QCD. Using the gauge/gravity duality, holographic QCD is described as 1+4 dimensional U($N_f$) gauge theory in flavor space with a background gravity, and its instanton solutions correspond to baryons. First, using the Witten Ansatz, we reduce holographic QCD into a 1+2 dimensional Abelian Higgs theory in a curved space and consider its topological aspect. We numerically obtain the Abrikosov vortex solution and investigate single baryon properties. Second, we study a single meron and two merons in holographic QCD. The single meron carrying a half-integer baryon number is found to have an infinite energy also in holographic QCD. We propose a new-type baryon excitation of the two-merons oscillation in the extra-direction of holographic QCD. 

\end{abstract}

\section{Introduction}

Quantum chromodynamics (QCD) has been established as the fundamental theory of the strong interaction. In spite of great success of perturbative QCD in high-energy regions, QCD exhibits strong-coupling nature beyond the perturbation in low-energy regions. Still now, to understand nonperturbative properties of QCD is one of the most important and difficult problems remaining in theoretical physics. 
As an interesting effective theory of nonperturbative QCD, holographic QCD was formulated using D-branes \cite{P95} and the gauge/gravity duality \cite{M98} 
in the superstring theory, and has been developed 
for the Yang-Mills theory \cite{W98} and massless 
QCD \cite{SS05,NSK07,HSSY07,HRYY07,HSS08,NSK09,CI12,BS14,RSR14,MS17}.

The superstring theory, which is well-defined in ten-dimensional space-time, has D$p$-branes \cite{P95} as ($p$+1)-dimensional massive soliton-like objects of fundamental strings. 
On the surface of $N$ D-branes, there exists a U($N$) gauge symmetry, and U($N$) gauge fields appear from open strings. Around the massive $N$ D-branes, 
a supergravity field proportional to $N$ is created in ten-dimensional space-time.
Since the gravity field depends on the distance from the D-brane, 
one more coordinate-dependence at least appears in the gravity side, 
which is called holography.

The D-brane generally leads to a SUSY system, reflecting superstring nature. 
In 1998, Witten \cite{W98} constructed non-SUSY gauge theories in this framework 
by imposing explicit SUSY breaking through $S^1$-compactification of D-branes 
with the periodic/anti-periodic boundary condition for bosons/fermions, 
in a similar manner to thermal SUSY breaking. 
The inverse radius of the $S^1$ is called as the Kaluza-Klein mass $M_{\rm KK}$, 
and the fermionic gaugino mass becomes $O(M_{\rm KK})$. 
Then, non-SUSY gauge theories are formed on the compactified D-branes 
at larger scale than 1/$M_{\rm KK}$. 

Actually, a four-dimensional non-SUSY U($N_c$) Yang-Mills theory is 
realized on $S^1$-compactified $N_c$ D4-branes, 
where only bosonic gauge fields ${\cal A}^\mu$ remain to be massless. 
On the other hand, the effect from $N_c$ D4-branes 
can be also described by the gravity field around it, 
under hypothesis of the gauge/gravity duality \cite{M98}.
In fact, the Yang-Mills theory or the gauge sector of QCD can be 
constructed with the $S^1$-compactified $N_c$ D4-branes, 
and it is transferred into 
a higher-dimensional gravity theory in the holographic framework. 
Owing to the strong-weak coupling duality, 
large-$N_c$ strong-coupling QCD is described 
with the weak-coupling gravity \cite{M98,W98}, 
and nonperturbative quantities of infrared QCD is calculable with 
the higher-dimensional classical gravitational theory.

\section{Construction of holographic QCD: the Sakai-Sugimoto model}

In 2005, Sakai and Sugimoto \cite{SS05} constructed 
four-dimensional massless QCD 
using the D4/D8/$\overline{\rm D8}$ multi-D-brane system, 
which consists of spatially $S^1$-compactified $N_c$ D4-branes 
attached with $N_f$ D8-$\overline{\rm D8}$ pairs, 
as shown in Fig.1(a). 
\begin{figure}[b]
\begin{center}
\includegraphics[width=7cm]{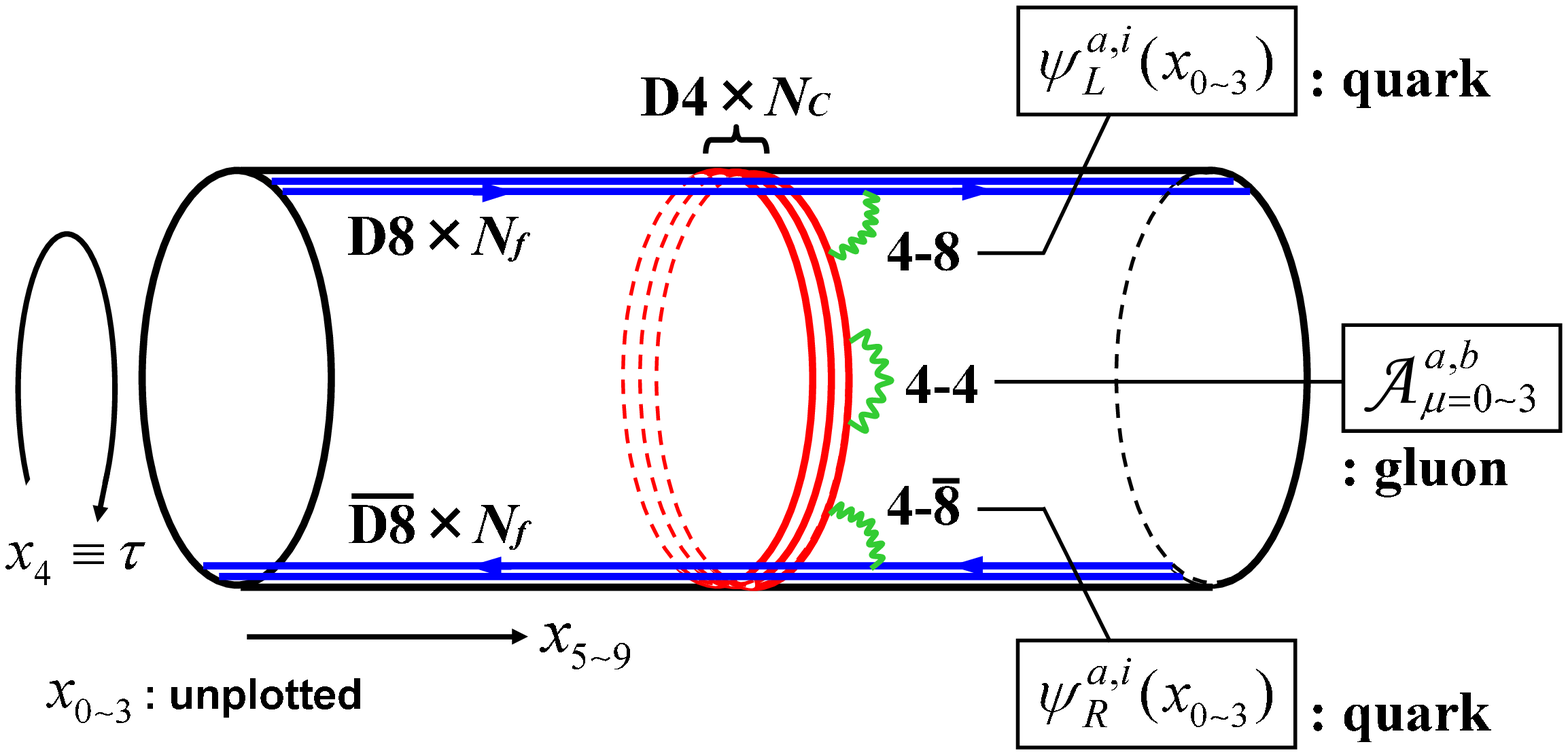}\\
\vspace{0.5cm}
\includegraphics[width=7cm]{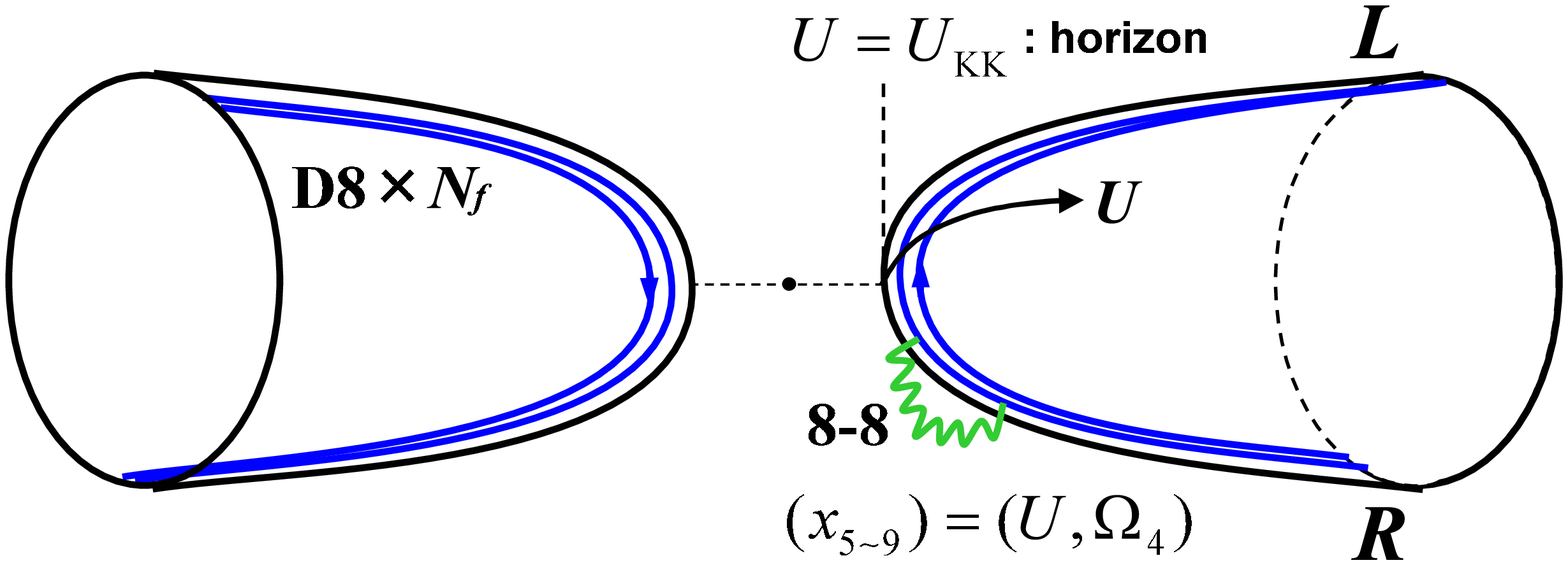}
\caption{The Sakai-Sugimoto model. 
(a) The upper figure shows the multi-D-brane configuration corresponding to massless QCD: 
spatially $S^1$-compactified $N_c$ D4-branes attached with 
$N_f$ D8-$\overline{\rm D8}$ pairs. 
Gluons appear from 4-4 strings, and 
quarks appear from 4-8 and 4-$\bar{8}$ strings. 
(b) The lower figure shows the D8-brane with D4 background gravity, which is generated 
from $N_c$ D4-branes as a result of the gauge/gravity duality. Mesons appear from 8-8 strings.}
\end{center}
\end{figure}

In holographic QCD, ``color'' and ``flavor'' are described 
as different physical objects, i.e., different D-branes: 
the D4 gives color and the D8 gives flavor. 
Here, gluons ${\cal A}^\mu \in {\rm su}(N_c)$ appear from 4-4 strings on the D4-brane, 
and (anti)quarks, which have both color and flavor, 
appear from 4-8 (4-$\bar{8}$) strings at the cross point 
between D4 and D8 ($\overline{\rm D8})$.
This D-brane system has an SU($N_c$) gauge symmetry and the exact chiral symmetry, 
and leads to four-dimensional QCD in the chiral limit \cite{SS05} 
at the larger scale than $M_{\rm KK}^{-1}$, i.e., the radius of the $S^1$.
In fact, at the infrared scale, this multi-D-brane system is described by 
1+3 dimensional massless QCD: 
\begin{eqnarray}
 {\cal L} = -\frac{1}{2} {\rm tr} \left( {\cal F}_{\mu\nu} {\cal F}^{\mu\nu} \right) 
+ \bar{q} i\gamma_\mu {\cal D}^\mu q.
\end{eqnarray}
Here, ${\cal F}^{\mu\nu}$ and ${\cal D}^\mu$ are the field strength and 
the covariant derivative operating in the color space, 
\begin{eqnarray}
{\cal F}^{\mu\nu} \equiv \frac{1}{ig}\left[{\cal D}^\mu, {\cal D}^\nu\right], 
\ \ {\cal D}^\mu \equiv \partial^\mu + ig{\cal A}^\mu,
\end{eqnarray}
with the QCD gauge coupling $g$.

In large-$N_c$ argument, 
$N_c$ D4-branes have a large mass proportional to $N_c$, 
which gives a large gravitational source.
Using the gauge/gravity duality, $N_c$ D4-branes can be replaced by a bulk gravity field, 
and the system becomes $N_f$ D8-branes in the presence 
of the background gravity of the $N_c$ D4-branes, as shown in Fig.1(b).
In large $N_c$ ($\gg N_f$), 
the gravitational contribution from D8/$\overline{\rm D8}$ can be neglected, 
which corresponds to the quenched approximation \cite{NSK07}.

In this framework, 
large 't~Hooft's coupling $\lambda \equiv g^2 N_c$ is taken, 
which means strong-coupling QCD and a weak-coupling gravity, 
due to the strong-weak duality. 
In large $N_c$ and large 't~Hooft's coupling $\lambda$, 
the $N_f$ D8-brane system in the D4 background gravity can be expressed with 
the Dirac-Born-Infeld (DBI) action in nine-dimensional space-time \cite{P95,SS05}, 
\begin{eqnarray}
S^{\rm DBI}_{\rm D8} = T_8 \int d^9x~ e^{-\phi}
{\rm tr}\bigl\{[-{\rm det}(g_{MN}+2\pi \alpha' F_{MN})]^{1/2}\bigr\},
\end{eqnarray}
where $F_{MN}$ is the field strength in the U($N_f$) flavor space on the D8 brane, 
and the U($N_f$) gauge field originates from 8-8 strings. 
$T_8$, $\phi$ and $\alpha'$ are quantities defined in the superstring theory \cite{SS05}.
Here, flavor degrees of freedom only remains, since 
the D4-brane with color is already replaced by the background gravity of $g_{MN}$.

This system possesses an SO(5) rotational symmetry in nine-dimensional space-time, 
and hence the DBI action is reduced to be five-dimensional.
In the leading of large $N_c$ and $\lambda$, the DBI action eventually becomes 
a five-dimensional U($N_f$) flavored Yang-Mills theory \cite{SS05} 
with a curved metric on an extra fifth-coordinate $w$, which 
corresponds to the distance from the D4-brane. 

In this way, from the multi-D-brane system which is infrared equivalent to massless QCD, 
one derives 1+4 dimensional Yang-Mills theory 
on U($N_f$) $\simeq {\rm SU}(N_f) \times$ U(1) 
in the leading order of $1/N_c$ and $1/\lambda$ expansions: 
\begin{eqnarray}
S_{\rm 5YM} &=& S_{\rm 5YM}^{{\rm SU}(N_f)} +S_{\rm 5YM}^{\rm U(1)}
\nonumber \\
&=& -\kappa \int d^4x dw \ \textrm{tr} 
\left[ \frac{1}{2}h(w)F_{\mu\nu} F^{\mu\nu} + k(w)F_{\mu w} F^{\mu w} \right]
\nonumber \\
&&-\frac{\kappa}{2} \int d^4x dw \ \textrm{tr} 
\left[ \frac{1}{2}h(w) \hat F_{\mu\nu} \hat F^{\mu\nu} 
+ k(w) \hat F_{\mu w} \hat F^{\mu w} \right].
\label{eq:5YM}
\end{eqnarray}
For $M, N = t, x, y, z, w$, the field strength is given by 
\begin{eqnarray}
F_{MN} &\equiv& \partial_M A_N-\partial_N A_M+i[A_M, A_N], 
\nonumber \\
\hat F_{MN} &\equiv& \partial_M \hat A_N-\partial_N \hat A_M 
\end{eqnarray}
with the five-dimensional SU($N_f$) gauge field $A^M(x^\mu,w)$ 
and U(1) gauge field $\hat A^M(x^\mu,w)$. 
Here, we take the $M_{\rm KK}=1$ unit and $\kappa = \frac{\lambda N_c}{216\pi^3}$. 
In Eq.(\ref{eq:5YM}), 
SU($N_f$) variables $A$ and U(1) variables $\hat A$ 
are completely separated, 
and hence $S_{\rm 5YM}$ is divided into the SU($N_f$) sector $S_{\rm 5YM}^{{\rm SU}(N_f)}$ 
and the U(1) sector $S_{\rm 5YM}^{{\rm U}(1)}$. 
Note that there appears background gravities $k(w)$ and $h(w)$ 
depending on the extra fifth-coordinate $w$, 
\begin{eqnarray}
k(w) = 1+w^2, \ \ h(w) = k(w)^{-1/3},
\end{eqnarray}
as a relic of $N_c$ D4-branes.

The $1/N_c$-leading holographic QCD 
has also the Chern-Simons (CS) term \cite{SS05,HSSY07}
as the next leading order of $1/\lambda$.
The CS term is a topological term responsible to anomalies in QCD, and 
its explicit form is 
\begin{eqnarray}
S_{\rm CS} &=& \frac{N_c}{24\pi^2}\int \left[ \frac{3}{2} \hat{A}{\rm tr}F^2 + \frac{1}{4}\hat{A}\hat{F}^2 +({\rm total~derivatives}) \right] \nonumber \\
&=& \frac{N_C}{24\pi^2}\epsilon_{MNPQ}\int d^4x dw [ \frac{3}{8}\hat{A}_0 {\rm tr}F_{MN}F_{PQ} - \frac{3}{2}\hat{A}_M{\rm tr}(\partial_0A_NF_{PQ}) \nonumber \\ 
&&+ \frac{3}{4}\hat{F}_{MN}{\rm tr}(A_0F_{PQ}) + \frac{1}{16}\hat{A}_0\hat{F}_{MN}\hat{F}_{PQ} - \frac{1}{4}\hat{A}_M \hat{F}_{0N}\hat{F}_{PQ}],
\label{eq:CS}
\end{eqnarray}
where $A$ is the SU($N_f$) gauge field and $\hat A$ the U(1) gauge field. 
Here and hereafter, the capital-letter index denotes the spatial index as $M=x, y, z, w$. 
In Eq.(\ref{eq:CS}), SU($N_f$) variables $A$ and U(1) variables $\hat A$ 
are dynamically mixed in $S_{\rm CS}$. 

To summarize, in the Sakai-Sugimoto model, 
holographic QCD is expressed by $S_{\rm HQCD}=S_{\rm 5YM}+S_{\rm CS}$,
up to the $1/N_c$ leading and the $1/\lambda$ next leading. 
Note that holographic QCD has a mathematical connection to infrared QCD 
through D-branes, and is successful 
to explain many phenomenological laws in hadron physics \cite{SS05}:
the vector-meson dominance (VMD), 
the Kawarabayashi-Suzuki-Riazuddin-Fayyazuddin (KSRF) relation, 
the Gell-Mann-Sharp-Wagner (GSW) relation, 
the hidden local-symmetry (HLS) picture, 
the Skyrme chiral-soliton picture \cite {S61,W79,ANW83} and so on.

\section{Baryons from instantons in holographic QCD}

In holographic QCD, 
baryons appear from instantons in the flavor space \cite{SS05,HSSY07},
and eventually they are described as chiral solitons \cite{NSK07}. 
Note here that holographic QCD already includes four-dimensional spatial coordinates 
$(x,y,z,w)$ including the extra fifth-coordinate $w$, and 
instantons can be naturally introduced in holographic QCD 
without necessity of the Euclidean process or the Wick rotation.

Actually, holographic QCD has instantons as topological solitons 
corresponding to the nontrivial homotopy group $\Pi_3({\rm SU}(N_f))={\bf Z}$ 
in spatial coordinates $(x,y,z,w)$ \cite{HSSY07,HRYY07}, 
and the baryon number $B$ is expressed by the Pontryagin index \cite{W98b}, 
\begin{eqnarray}
B = \frac{1}{16\pi^2} \int d^3x dw \ {\rm tr}\bigl( F_{MN} \tilde{F}_{MN} \bigr). 
\end{eqnarray}

After the mode expansion along the extra $w$-direction, 
holographic QCD becomes the 1+3 dimensional Skyrme model 
including vector and axial-vector mesons \cite{SS05}, and baryons eventually 
become topological hedgehog solitons corresponding to the nontrivial homotopy group 
$\Pi_3({\rm SU}(N_f)_L \times {\rm SU}(N_f)_R/{\rm SU}(N_f)_V)={\bf Z}$. 

In 2006, we studied baryons as brane-induced Skyrmions in holographic QCD for $N_f =2$ 
in the leading of $1/N_c$ and $1/\lambda$ \cite{NSK07}, where 
we have included the background gravity precisely, 
but considered only light-mesons (pions and $\rho$-mesons) contributions.
In 2007, Hata et al. described baryons from instantons in holographic QCD 
for $N_f =2$ \cite{HSSY07}, but they neglected the background gravity 
$h(w)$ and $k(w)$, because of the difficulty to deal with instantons in the curved space.
In 2014, Bolognesi et al. numerically investigated 
the holographic baryon, precisely keeping the background gravity, 
using the SO(3) rotational symmetry \cite{BS14,RSR14}.

\subsection{Witten Ansatz for holographic QCD}

To begin with, in a similar manner in Refs.\cite{BS14,RSR14}~, 
we investigate baryons from instantons 
in the curved space with the background gravity $h(w)$ and $k(w)$ 
in the framework of 1+4 dimensional holographic QCD for $N_f =2$.
The strategy is to use the Witten Ansatz \cite{W77}, 
which keeps the spatially rotational SO(3) symmetry 
on ${\bf x} \equiv (x, y, z)$ and ${\bf A} \equiv (A_x, A_y, A_z)$.

For the SU($N_f$=2) gauge field $A=A^a\frac{\tau^a}{2} \in {\rm su}(2)_f$ 
in holographic QCD, we take the Witten Ansatz \cite{W77}:
\begin{eqnarray}
A_0^a(t,x,y,z,w) &=& a_0 (t,r,w) \hat{x}^a, \nonumber \\
A_i^a(t,x,y,z,w) &=& \frac{\phi_2(t,r,w)+1}{r}\epsilon_{iak}\hat{x}_k + \frac{\phi_1(t,r,w)}{r} \hat{\delta}_{ia} + a_r(t,r,w) \hat{x}_i\hat{x}_a, \nonumber \\
A_w^a(t,x,y,z,w) &=& a_w (t,r,w) \hat{x}^a, 
\end{eqnarray}
with $r \equiv (x_i x_i)^{1/2}$, $\hat x_i \equiv x_i/r$
and $\hat \delta_{ij} \equiv \delta_{ij}-\hat x_i \hat x_j$.
Note that this form has the SO(3) symmetry 
on the spatial rotation of ${\bf x} \rightarrow R {\bf x}$ and ${\bf A} \rightarrow R {\bf A}$, 
together with the isospin rotation $\vec \tau \rightarrow R \vec \tau$, 
for arbitrary SO(3) matrix $R$. 
This symmetry keeps the hedgehog structure $\tau^a \hat x^a$ invariant, 
and is useful in describing topological solitons.


By way of the Witten Ansatz, 1+4 dimensional SU(2)$_f$ holographic QCD is reduced into 
a 1+2 dimensional Abelian Higgs theory on a curved space.
In fact, the leading term $S_{\rm 5YM}^{{\rm SU(2)}_f}$ is rewritten as 
\begin{eqnarray}
S_{\rm 5YM}^{{\rm SU(2)}_f} &=& -\kappa \int d^4xdw {\rm tr} \left[ 
\frac{1}{2} h(w) F_{\mu\nu} F^{\mu\nu} + k(w)F_{\mu w} F^{\mu w} \right] \nonumber \\
&=& 4\pi\kappa \int^\infty_{-\infty} dt \int^\infty_0 dr \int^\infty_{-\infty} dw 
\biggl[ h(w) (|D_0\phi|^2 - |D_1\phi|^2) - k(w) |D_2\phi|^2 \nonumber \\
&&- \frac{h(w)}{2r^2}(1-|\phi|^2)^2 
+ \frac{r^2}{2} \{ h(w) f_{01}^2 + k(w) f_{02}^2 - k(w) f_{12}^2 \} \biggr],
\label{eq:S5YM}
\end{eqnarray}
where the Abelian Higgs field $\phi(t,r,w) \in {\bf C}$,
the Abelian gauge field $a_\mu(t,r,w)$, the covariant derivative $D_\mu$ 
and the field strength $f_{\mu\nu}$ in the Abelian Higgs theory are 
\begin{eqnarray}
\phi \equiv \phi_1+i\phi_2, \
a_\mu \equiv (a_0,a_r,a_w), \ 
D_\mu \equiv \partial_\mu - i a_\mu, \
f_{\mu\nu} \equiv \partial_\mu a_\nu-\partial_\nu a_\mu. \ \
\end{eqnarray}
Here, we have used $(0,1,2)=(t,r,w)$ for the index of 1+2 dimensional coordinates. 

\subsection{Vortex description of baryon}

Here, we investigate the topological correspondence between the above two theories, 
by considering the topological charge corresponding to the baryon number $B$. 
In holographic QCD, instantons are responsible to the baryon number \cite{SS05,HSSY07}. 
Remarkably, the instanton in holographic QCD is described 
as the Abrikosov vortex in the 1+2 dimensional Abelian Higgs theory \cite{W77},
as is explicitly shown below.

For the SU(2)$_f$ configuration in the Witten Ansatz, 
the topological density $\rho_B$ in $(x,y,z,w)$-space is given by
\begin{eqnarray}
\rho_B &\equiv& \frac{1}{16\pi^2} {\rm tr}\bigl( F_{MN} \tilde{F}_{MN} \bigr) 
= \frac{1}{32\pi^2} \epsilon_{MNPQ} {\rm tr}\bigl( F_{MN} F_{PQ} \bigr)
\nonumber \\
&=&\frac{1}{8\pi^2r^2}\epsilon_{ij}\partial_i
\{a_j(1-|\phi|^2)+\partial_j \theta \cdot |\phi|^2\},
\end{eqnarray}
where $\theta\equiv {\rm arg} \phi$ and 
the Roman indices $i, j$ take $(1,2)=(r,w)$.
The topological density $\rho_B$ is expressed as a total derivative. 
Note that $\rho_B$ is independent from the spatial direction $(\hat x, \hat y, \hat z)$, 
and thus takes an SO(3) rotationally symmetric form of $\rho_B(r,w)$.

Then, the baryon number $B$ or the Pontryagin index is written by 
a contour integral in $(r,w)$-plane, 
\begin{eqnarray}
B &=& \int d^3x dw~ \rho_B =\frac{1}{2\pi}
\int_0^\infty dr \int_{-\infty}^\infty dw
\epsilon_{ij}\partial_i
\{a_j(1-|\phi|^2)+\partial_j \theta \cdot |\phi|^2\}
\nonumber \\
&=& \oint_{r \ge 0} d{\bf s} \cdot \{{\bf a}(1-|\phi|^2)+\nabla \theta \cdot |\phi|^2\}
= \oint_{r \ge 0} d{\bf s} \cdot \nabla \theta, 
\label{eq:bnum}
\end{eqnarray}
where $\oint_{r \ge 0}$ denotes the contour integral around 
the whole half-plane of $(r,w)$ with $r\ge 0$. 
Here, we have used $|\phi(r=0, ^{\forall} \! w)|=1$ and 
$|\phi(\infty)|=1$ in $(r,w)$-plane, as the boundary condition 
for the finite-energy solution. (See Eqs.(\ref{eq:S5YM}) and (\ref{eq:E5YM}).)
Thus, the baryon number $B$ is converted into the vortex number in this formalism.

\subsection{Baryons in leading order of $1/N_c$ and $1/\lambda$}

In the reduced low-dimensional Abelian Higgs theory, 
the energy of the static configuration is simply expressed 
in the temporal gauge $a_0=0$ by 
\begin{eqnarray}
E_{\rm 5YM}^{{\rm SU(2)}_f} &=& 4\pi\kappa \int^\infty_0 dr \int^\infty_{-\infty} dw 
\biggl[ h(w) |D_1\phi|^2 + k(w) |D_2\phi|^2 
\nonumber \\
&&+ \frac{h(w)}{2r^2}(1-|\phi|^2)^2 + \frac{r^2}{2} k(w) f_{12}^2 \biggr].
\label{eq:E5YM}
\end{eqnarray}

In the leading order of $1/N_c$ and $1/\lambda$, one only has to consider 
the main SU($N_f$) part, $S_{\rm 5YM}^{{\rm SU(2)}_f} $ 
and $E_{\rm 5YM}^{{\rm SU(2)}_f}$, 
in holographic QCD, since the U(1) part is completely separated 
in $S_{\rm 5YM}$ in Eq.(\ref{eq:5YM}). 
Furthermore, the classical-level analysis is enough meaningful, 
because this theory is weak-coupling corresponding to strong-coupling QCD 
via the strong-weak duality. 
Nawa et al. \cite{NSK07} investigated baryons as the brane-induced Skyrmions 
by truncating the higher mesonic modes than $M_{\rm KK}$, 
and obtained finite-size baryons from the DBI action $S_{\rm 5YM}$.
However, if no truncation is done for the mesonic modes in $S_{\rm 5YM}$, 
the instanton solution or the ground-state baryon shrinks 
to be point-like \cite{HSSY07}, which means that the derivative 
becomes quite large and the low-energy treatment is to be corrected. 
As a remedy, Hata et al. \cite{HSSY07,HRYY07} included the Chern-Simons (CS) term,
which is the next leading order of the $1/\lambda$ expansion, 
and obtained a finite-size instanton for the baryon.

Here, in the leading order of $1/N_c$ and $1/\lambda$, 
we numerically calculate the ground-state baryon in Eq.(\ref{eq:E5YM}), 
and also find 
the shrinking of the Abrikosov vortex solution into the point $(r,w)=(0,0)$, 
which corresponds to a point-like instanton at $w=0$ \cite{W77}.

\subsection{U(1) sector in holographic QCD}

In the leading order of $1/N_c$ and $1/\lambda$, 
the U(1) sector is decoupled with the SU($N_f$) sector, 
and has no physical importance.
At the next leading order of $1/\lambda$, however, 
the U(1) sector is coupled with the SU($N_f$) sector through the CS term,
and has physical importance \cite{HSSY07}. 

Also for the U(1) gauge field $\hat A$, 
we respect the spatial SO(3) rotational symmetry \cite{BS14,RSR14}
as in the Witten Ansatz, and impose 
\begin{eqnarray}
\hat A_i(t,x,y,z,w) &=& \hat a_r(t,r,w) \hat{x}_i, 
\end{eqnarray}
while $\hat A_0$ and $\hat A_w$ are treated to be arbitrary.
In this case, one finds $\hat F_{ij}=0$ and can take the $\hat a_r=0$ gauge.
Then, the $1/\lambda$-leading term $S_{\rm 5YM}^{\rm U(1)}$ 
is written as 
\begin{eqnarray}
S_{\rm 5YM}^{\rm U(1)}
&=& 
\frac{\kappa}{2}
\int d^4x dw \{ h(w) \hat F_{0i}^2+k(w) \hat F_{0w}^2- k(w) \hat F_{iw}^2\} 
\nonumber \\
&=& \int d^4x dw \biggl[ 
\frac{1}{2}\hat A_0 K \hat A_0-\frac{\kappa}{2} k(w) (\partial_i \hat A_w)^2
\biggr],
\end{eqnarray}
using the SO(3)-symmetric non-negative hermite kernel 
\begin{eqnarray}
K \equiv -\kappa \{ h(w) \partial_i^2 + \partial_w k(w) \partial_w \}
=-\kappa \{ h(w) \frac{1}{r^2}\partial_r r^2 \partial_r 
+ \partial_w k(w) \partial_w \}.
\end{eqnarray}

\subsection{Baryons up to next leading order of $1/\lambda$}

Next, let us consider the CS term $S_{\rm CS}$ 
as the next leading order of the $1/\lambda$ expansion. 
For the static SO(3)-rotationally symmetric configuration in the $A_0=0$ gauge, 
the CS term $S_{\rm CS}$ in Eq.~(\ref{eq:CS}) 
is simply expressed as \cite{HSSY07,BS14,RSR14}
\begin{eqnarray}
S_{\rm CS} = \frac{N_c}{2}\int d^4x dw~ \rho_B \hat A_0,
\end{eqnarray}
which is the Coulomb-type interaction between 
the U(1) gauge potential $\hat A_0$ and the topological density 
$\rho_B \equiv \frac{1}{16\pi^2} {\rm tr}(F_{MN} \tilde{F}_{MN})$.

Then, the total U(1) action 
depending on the U(1) gauge field $\hat A$ is written as 
\begin{eqnarray}
S^{\rm U(1)} &\equiv& S_{\rm 5YM}^{\rm U(1)}+S_{\rm CS} 
\nonumber \\
&=& \int d^4x dw \biggl[ 
\frac{1}{2}\hat A_0 K \hat A_0 +\frac{N_c}{2}\rho_B \hat A_0 
-\frac{\kappa}{2} k(w) (\partial_i \hat A_w)^2
\biggr],
\end{eqnarray}
which leads to the field equations,
\begin{eqnarray}
K \hat A_0 +\frac{N_c}{2}\rho_B=0, \ \ \partial_i^2 \hat A_w=0.
\end{eqnarray}
Note again that the classical-level analysis is enough meaningful, 
because of the weak coupling corresponding to strong-coupling QCD.

Thus, for the static solution, 
we obtain from $S^{\rm U(1)}$ the additional energy term 
\begin{eqnarray}
E^{\rm U(1)} &=&\frac{N_c^2}{8}\int d^3 x dw~ \rho_B K^{-1} \rho_B
\nonumber \\
&=&\frac{N_c^2}{8}\int d^3 x dw \int d^3 x' dw' 
\rho_B(\vec x,w) K^{-1}(\vec x,w; \vec x',w') \rho_B(\vec x',w'),
\end{eqnarray}
apart from the leading energy term $E_{\rm 5YM}^{{\rm SU(2)}_f}$.
This additional term $E^{\rm U(1)}$ is the repulsive Coulomb-type energy 
among the topological density $\rho_B$, and this repulsive force prevents 
the spatial gathering of the topological-charge distribution \cite{HSSY07}.

Since the kernel $K$ and the topological density $\rho_B$ are 
SO(3)-rotationally symmetric, the additional energy is expressed 
only with the $(r,w)$-coordinates:
\begin{eqnarray}
E^{\rm U(1)} = 2\pi^2 N_c^2 \int_0^\infty dr \int_{-\infty}^\infty dw 
&&\int_0^\infty dr' \int_{-\infty}^\infty dw'
\nonumber  \\ 
&& \tilde \rho(r,w) \tilde K^{-1}(r,w; r',w') \tilde \rho(r',w'),
%
\end{eqnarray}
using $\tilde \rho(r,w) \equiv r^{2}\rho_B(r,w)$ and 
the hermite kernel $\tilde K$ in $(r,w)$-space, 
\begin{eqnarray}
\tilde K \equiv 4\pi r^2 K
=-4\pi \kappa \{h(w) \partial_r r^2 \partial_r + r^2 \partial_wk(w) \partial_w\}.
\end{eqnarray}

For the numerical calculation, we consider the total energy 
$E \equiv E_{\rm 5YM}^{{\rm SU(2)}_f} +E^{\rm U(1)}$ 
corresponding to $S_{\rm HQCD}=S_{\rm 5YM}+S_{\rm CS}$, 
and obtain the $B=1$ vortex solution on lattice-discretized $(r,w)$-plane.
Owing to the SO(3) rotational symmetry, 
the numerical cost is significantly reduced 
from (4D volume)$^2$ to (2D volume)$^2$.

\subsection{Abrikosov vortex solution for baryon in holographic QCD}

Now, we show the numerical result 
for the vortex solution for the baryon with $B=1$. 
As for the two parameters $M_{\rm KK}$ and $\kappa$ in holographic QCD, 
we take $M_{\rm KK}\simeq$ 948MeV and $\kappa=7.46 \times10^{-3}$, 
to reproduce $f_\pi \simeq$ 92.4MeV and $m_\rho \simeq$ 776MeV \cite{SS05,NSK07}.

\begin{figure}[h]
\centerline{\includegraphics[width=13cm]{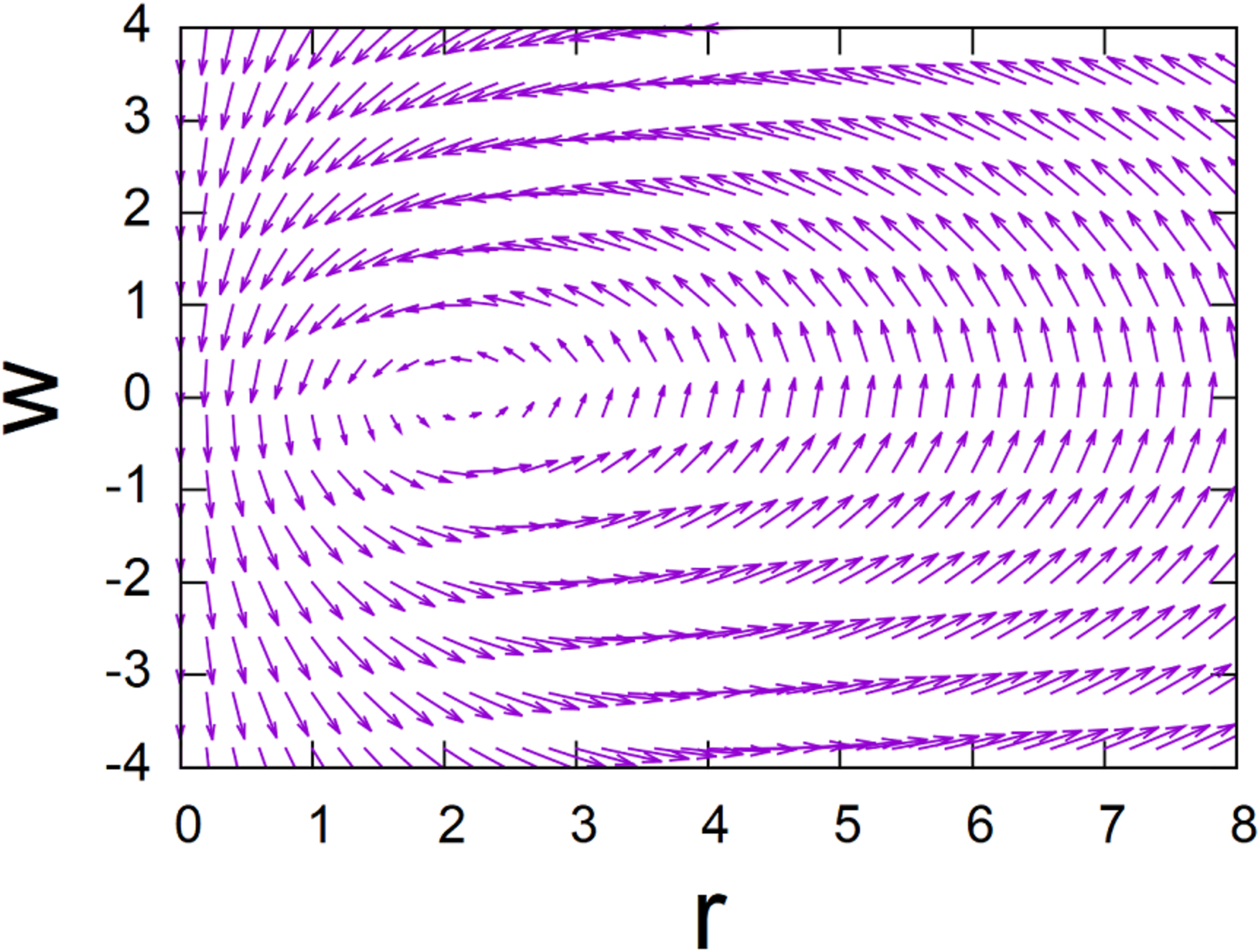}}
\centerline{\includegraphics[width=13cm]{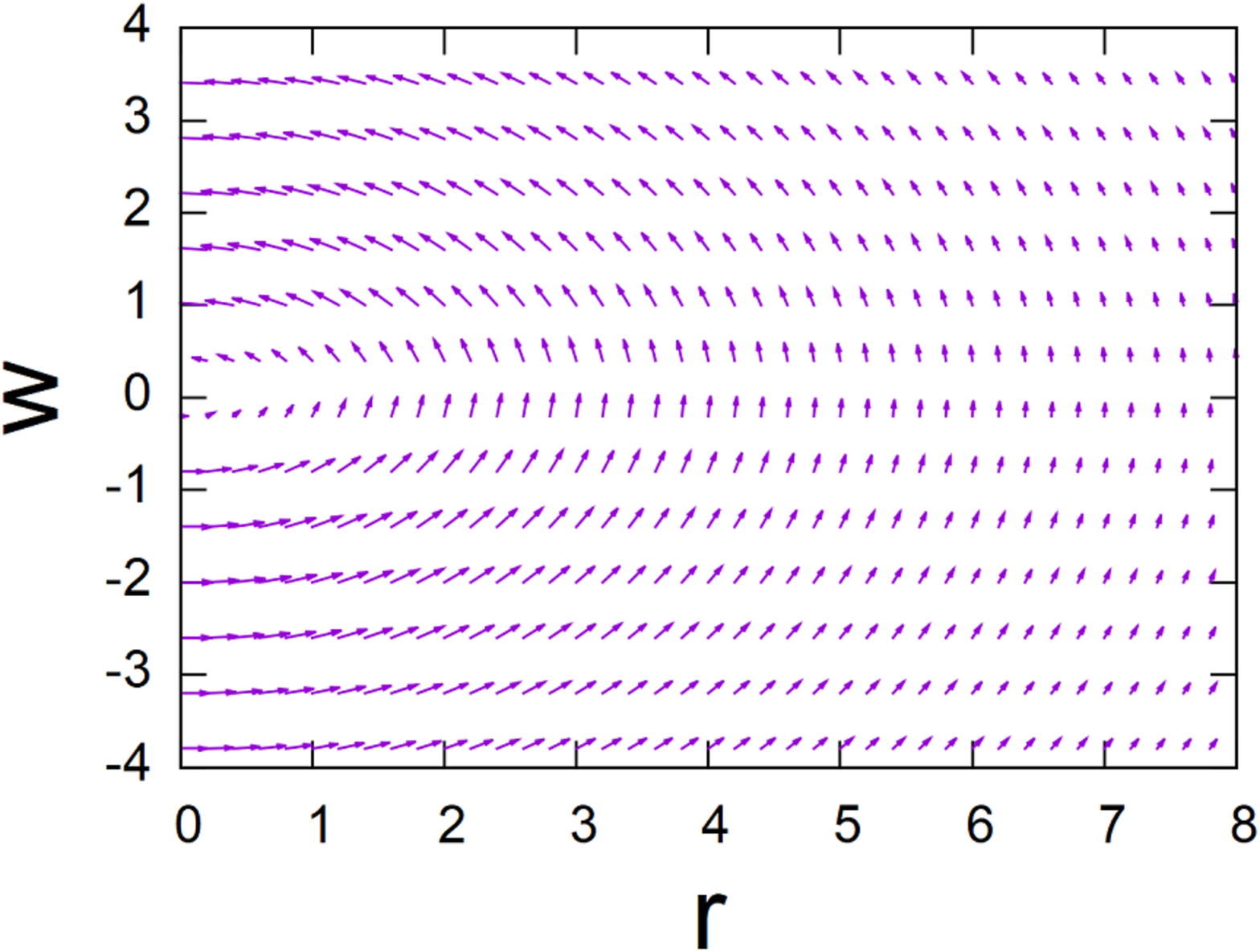}}
\caption{
The static Abrikosov vortex solution for the baryon with $B=1$ 
in the 1+2 Abelian Higgs theory reduced from 
1+4 holographic QCD using the Witten Ansatz.
The upper figure shows the Higgs-field configuration $\vec \phi=(\phi_1, \phi_2)$,
and the lower figure the Abelian gauge configuration $\vec a=(a_r, a_w)$ in $(r,w)$-plane 
in the Landau gauge. 
As the topological structure, the single winding number 
around $(r_0,w_0) \simeq (2,0)$, the zero point of the Higgs field $\vec \phi$, 
corresponds to the single baryon number $B=1$. 
The axis unit is $M_{\rm KK}^{-1} \simeq 0.2{\rm fm}$.}
\label{fig:vorh}
\end{figure}

Figure~\ref{fig:vorh} shows the Higgs-field configuration $\vec \phi=(\phi_1, \phi_2)$ 
and the Abelian gauge configuration $\vec a=(a_r, a_w)$ in $(r,w)$-plane 
for the static Abrikosov vortex solution 
in the 1+2 dimensional Abelian Higgs theory in the Landau gauge, 
$\partial_r a_r+\partial_w a_w=0$.

In Fig.\ref{fig:vorh}, the Higgs field $\vec \phi(r,w)$ has the zero point 
at $(r,w)=(r_0,w_0)\simeq (2,0)$, and forms the topological structure of 
the Abrikosov vortex around it.
In fact, we find the single winding number around the vortex center $(r_0,w_0)$, 
and this corresponds to the single baryon number $B=1$, 
as shown in Eq.(\ref{eq:bnum}). 

Since the $r$-coordinate of the vortex center 
corresponds to the instanton size in the Witten Ansatz \cite{W77}, 
from the vortex center $(r_0,w_0) \simeq (2,0)$, 
the instanton size $a$ is estimated as $a=r_0$, i.e., 
$a =r_0 M_{\rm KK}^{-1} 
\simeq 0.4{\rm fm}$ in the physical unit. 

\begin{figure}[h]
\hspace{2.05cm}
\includegraphics[width=5cm, angle=270]{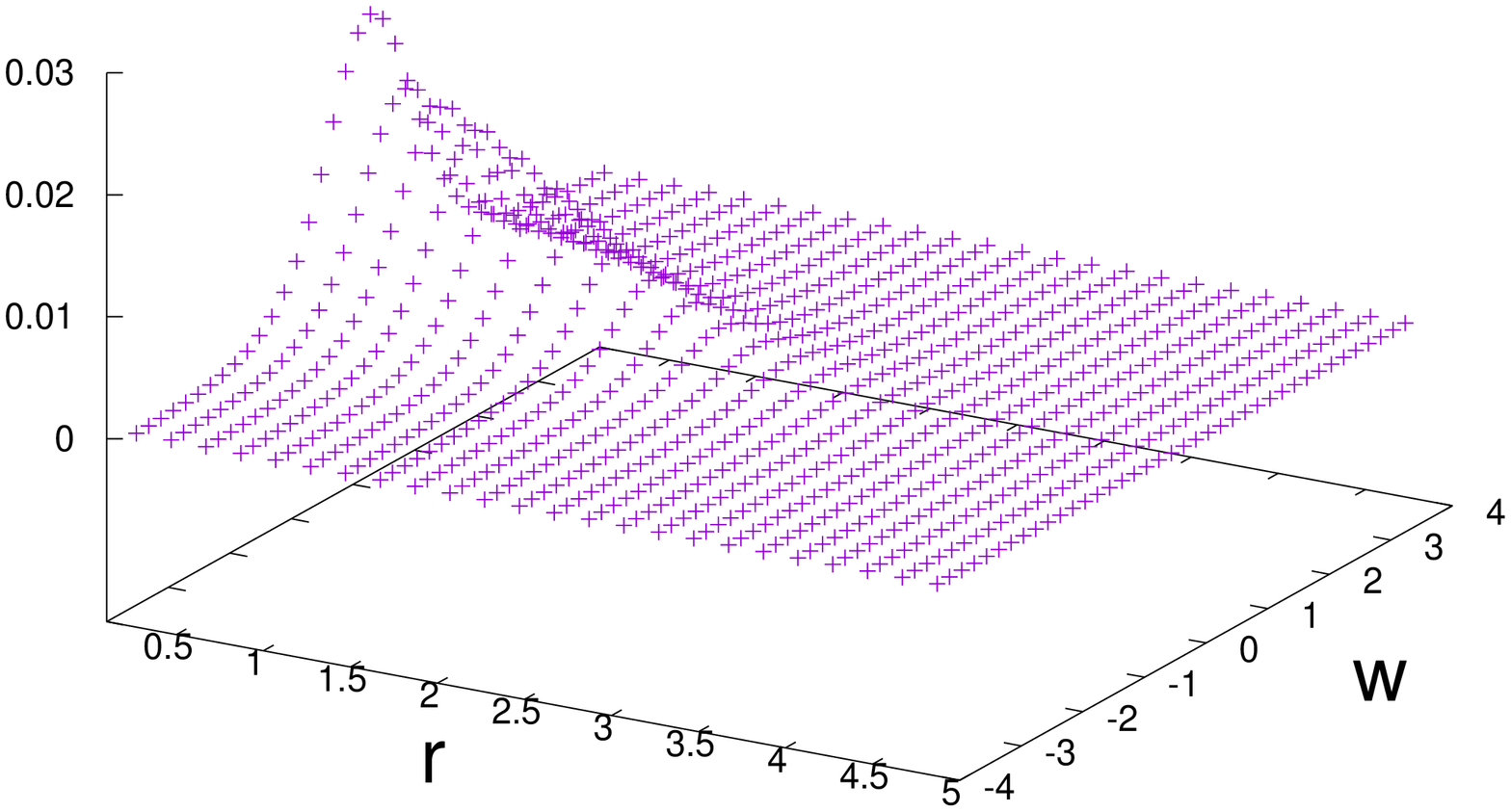}
\hspace{-0.5cm}
\includegraphics[width=5cm, angle=270]{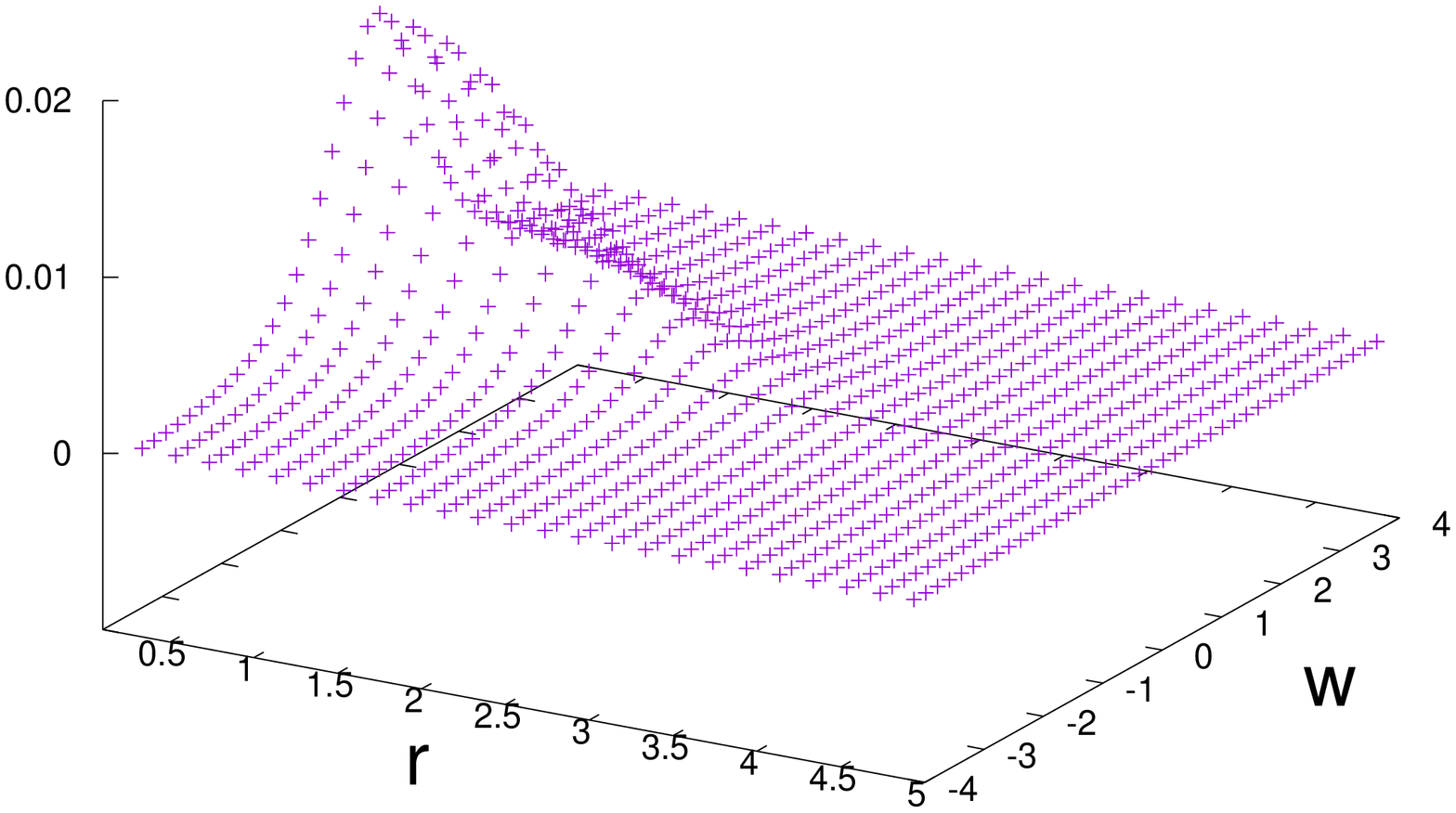}
\caption{
The left and right figures show the topological density $\rho_B(r,w)$ 
and the energy density ${\cal E}(r,w)$, respectively, 
in $(r,w)$-space for the Abrikosov vortex solution 
in the 1+2 Abelian Higgs theory, 
reduced from 1+4 holographic QCD.
The scale unit is $M_{\rm KK}=1$.
The integrated topological density is the baryon number $B=1$.
}
\label{fig:density}
\end{figure}

Figure~\ref{fig:density} shows 
the topological density $\rho_B(r,w)$ 
and the energy density ${\cal E}(r,w)$ 
in $(r,w)$-plane for the Abrikosov vortex solution. 
For both topological and energy densities, 
we find an extended lump around the vortex center. 
Here, the integrated topological density is the baryon number $B$, 
and the baryon mass $M_B$ is given by the integrated energy density 
as the static energy of the ground-state baryon: 
\begin{eqnarray}
B&=&\int d^3x dw~ \rho_B
=4\pi \int_0^\infty dr r^2 \int_{-\infty}^\infty dw~ \rho_B(r,w),
\\
M_B&=&\int d^3x dw~ {\cal E}(r,w)=
4\pi \int_0^\infty dr r^2 \int_{-\infty}^\infty dw~ {\cal E}(r,w).
\end{eqnarray}

\begin{figure}[h]
\centerline{\includegraphics[width=7cm, angle=270]{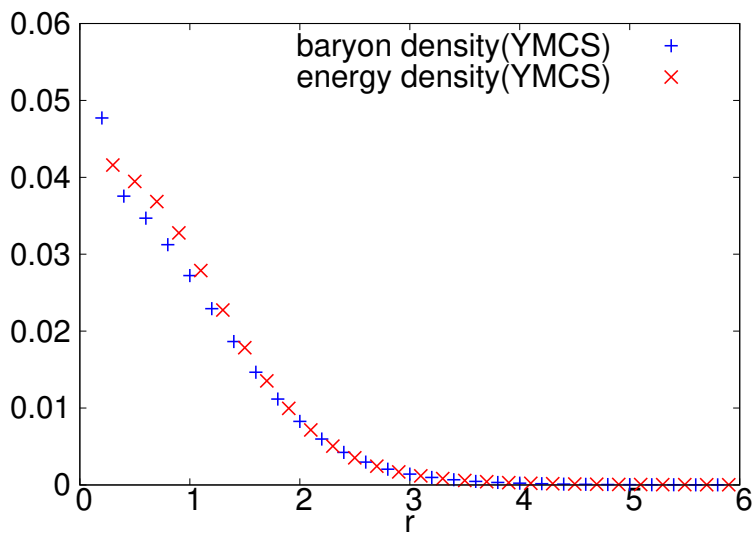}}
\caption{
The energy density ${\cal E}(r)$ and the baryon-number density $\rho_B(r)$ 
inside the baryon as the function of $r$ in holographic QCD. 
In this figure, the unit of $M_{\rm KK}=1$ is taken, and hence 
the physical unit of the horizontal axis is $M_{\rm KK}^{-1} \simeq 0.2{\rm fm}$.}
\label{fig:size}
\end{figure}

The ordinary energy and baryon-number densities in the three-dimensional space 
are obtained by the integration over the extra coordinate $w$ as
\begin{eqnarray}
{\cal E}(r) \equiv \int_{-\infty}^\infty dw~{\cal E}(r,w), \ \ 
\rho_B(r) \equiv \int_{-\infty}^\infty dw~\rho_B(r,w).
\end{eqnarray}
We show in Fig.~\ref{fig:size} the energy density 
${\cal E}(r)$ and the baryon-number density $\rho_B(r)$
inside the baryon as the function of $r$ in holographic QCD.

Thus, the baryon mass $M_B$ and the baryon size are estimated as
\begin{eqnarray}
M_B=\int d^3x ~{\cal E}(r) \simeq 998 {\rm MeV}, 
\ 
\sqrt{\langle r^2 \rangle} 
\equiv \sqrt{\frac{\int d^3x~ {\cal E}(r) r^2}{\int d^3x~ {\cal E}(r)}} 
\simeq 0.48{\rm fm}.
\end{eqnarray} 

In a similar manner in the Skyrme model \cite{ANW83}, 
to perform semi-classical quantization of the baryonic soliton 
might be interesting and important \cite{HSSY07,HSS08}, 
but such a mass correction from the rotational energy of the soliton 
is only $O(1/N_c)$ and is much higher order in the $1/N_c$ expansion, 
in comparison with the baryon mass of $O(N_c)$.

\section{Meron in holographic QCD and two-merons oscillation modes}

As a next new subject, we study a meron \cite{CDG7779}, 
which has a half topological charge, in holographic QCD for the first time.

The meron is a curious classical solution of the 
{\it Euclidean} four-dimensional Yang-Mills theory, 
since it is zero-size ``half-instanton'' with the half topological charge ($Q=1/2$)  
and has an infinite Euclidean action ($S=\infty$). 
However, a recent study reports that such a meronic configuration 
can have a finite energy in the 1+4 dimensional Yang-Mills theory 
in the presence of some background gravity like a black hole \cite{CGORS19}.

Then, we study meron configurations 
in $(x,y,z,w)$-space in 1+4 dimensional U($N_f$) holographic QCD, 
where the background gravity exists.
It is notable that, as well as instantons, 
the meron can be considered 
in {\it spatial} $(x,y,z,w)$-coordinate space 
in 1+4 dimensional U($N_f$) holographic QCD with $N_f \ge 2$  
even in the {\it Minkowski} metric, 
and such a meron configuration in holographic QCD 
corresponds to a half baryon-number object with $B=1/2$ in QCD.

Before proceeding merons in holographic QCD, 
we briefly summarize traditional merons 
in the four-dimensional Euclidean Yang-Mills theory, 
where the space-time is Euclidean and it is difficult to consider 
Minkowski-time evolution of merons. 

In the four-dimensional Euclidean SU(2) Yang-Mills theory,
the single meron is expressed by
\begin{eqnarray}
A_\mu^a=-\frac{1}{2}\eta_{\mu\nu}^a \frac{x_\nu}{x^2},
\end{eqnarray}
where $\eta_{\mu\nu}^a$ denotes the 't~Hooft symbol \cite{CDG7779}. 
For the meron, the topological charge $Q$ is one half:
\begin{eqnarray}
Q \equiv \frac{1}{16\pi^2} \int d^4x \ {\rm tr}\bigl( F_{\mu\nu} \tilde{F}_{\mu\nu} \bigr) 
=\frac{1}{2}.
\end{eqnarray}
On the other hand, the action of the single meron is infinite in the infinite volume 
\cite{CDG7779}, 
and gradually increases as $\ln L$ for the system with the size $L^4$:
\begin{eqnarray}
S_L \equiv \frac{1}{16\pi^2} \int^L d^4x \ {\rm tr}\bigl( F_{\mu\nu} F_{\mu\nu} \bigr) 
\sim 3\pi^2 \ln L.
\end{eqnarray}
It is notable that divergence of the meron action is logarithmic and weak \cite{CDG7779}.
Then, its divergence or finiteness is nontrivial in the presence of background gravity.

Now, we proceed merons in 1+4 dimensional holographic QCD.  
Note here that holographic QCD already has four spatial coordinates $(x,y,z,w)$ 
and merons can be introduced in this four-dimensional space, 
without necessity of the Euclidean process. 
Furthermore, we can consider the Minkowski-time ($t$) evolution of merons 
in 1+4 dimensional holographic QCD, 
unlike the four-dimensional Euclidean Yang-Mills theory.

\subsection{Single meron in holographic QCD}

To begin with, we consider a single meron 
in 1+4 dimensional U($N_f$) holographic QCD with $N_f=2$. 
First, we consider a smeared-type meron configuration \cite{CDG7779} 
in spatial coordinates $(x,y,z,w)$ in holographic QCD: 
\begin{eqnarray}
A_M^a=-\frac{1}{2}\eta_{MN}^a \frac{x_N}{x^2+a^2},
\end{eqnarray}
with the meron size $a$.
Here, the capital indices $M, N$ denote the spatial coordinates $(x,y,z,w)$, and 
the extra coordinate $w$ plays a role of Euclidean time.

For the meron configuration, we consider the energy 
at the leading of $1/N_c$ and $1/\lambda$: 
\begin{eqnarray}
E_{\rm 5YM}^{{\rm SU}(N_f)}= \kappa \int d^3x dw \ \textrm{tr} 
\left[ \frac{1}{2}h(w)F_{ij}^2 + k(w)F_{kw}^2 \right], 
\end{eqnarray}
where the first and second terms can be regarded 
as the ``magnetic energy'' and the ``electric energy'', respectively, 
when $w$ is regarded as a Euclidean time.

As the result, we find that the ``magnetic energy'' of the meron configuration 
becomes finite, owing to the reduction gravitational factor $h(w)$, 
but the ``electric energy'' of the meron is more strongly divergent, 
due to the increasing gravitational factor $k(w)$.
Even if one includes the CS term of the next leading of $1/\lambda$,
the total energy of the meron configuration is infinite, 
because the additional energy is repulsive and positive. 

Next, we numerically study the {\it single meron solution}  
and its energy in holographic QCD, using the technique in the previous section. 
In fact, we consider the SO(3) rotational-invariant meron configuration, 
where holographic QCD is generally reduced into 
1+2 dimensional Abelian Higgs theory. 
We numerically examine the single-meron solution 
with the half topological charge $Q=B=1/2$ in the Abelian Higgs theory 
i) at the leading order and ii) up to the next leading order of $1/\lambda$ 
including the CS term, respectively.
For both cases, we find that the energy of 
the single meron solution with $Q=1/2$ 
is divergent when the volume goes to infinite. 

In fact, there is no single meron with a finite energy in holographic QCD, 
which means natural absence of such a half baryon-number object with $B=1/2$ 
in QCD. 

\subsection{Meron-pair oscillation in holographic QCD}

\begin{figure}[h]
\vspace{-0.5cm}
\centerline{\includegraphics[width=7cm]{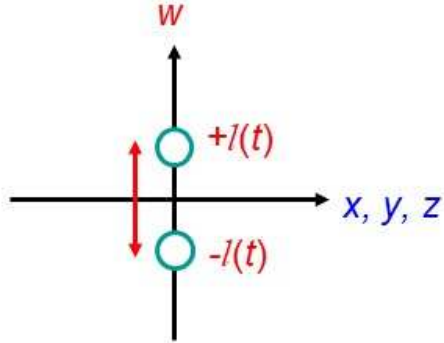}}
\vspace{-0.5cm}
\caption{
A schematic figure of a new-type baryon excitation 
of two smeared-merons oscillation 
in the extra fifth-direction in holographic QCD. 
The oscillation energy of the meron-pair is finite.
Since this oscillation is in the extra $w$-direction, 
which is independent of ordinary coordinates, $x,y,z$, 
this type of oscillating excitation is expected to appear for any baryon universally, 
as an interesting possibility.
}
\label{fig:merons}
\end{figure}

Finally, we study the two-merons configuration and its time-dependent oscillation 
in the extra fifth-direction in holographic QCD, 
as is schematically depicted in Fig.\ref{fig:merons}. 
The two-merons configuration is topologically identified as a kind of 
a single instanton, and the time-dependent oscillation 
can be regarded as an excitation of a single baryon in holographic QCD. 

We consider the two smeared merons with the same size $a$ 
and their centers being located at 
$x_1=(0,0,0,l(t))$ and $x_2=(0,0,0,-l(t))$ in the $(x,y,z,w)$-space,
\begin{eqnarray}
A_M^a
&=&-\frac{1}{2}\eta_{MN}^a 
\biggl[\frac{(x-x_1)_N}{(x-x_1)^2+a^2} + \frac{(x-x_2)_N}{(x-x_2)^2+a^2} \biggr]
\nonumber \\
&=&-\frac{1}{2}\eta_{MN}^a 
\biggl[\frac{\{x-l(t) \hat w\}_N}{\{x-l(t) \hat w\}^2+a^2} + \frac{\{x+l(t) \hat w\}_N}{\{x+l(t) \hat w\}^2+a^2} \biggr],
\end{eqnarray}
where $\hat w$ denotes the unit vector in the $w$-direction 
and $M, N$ take $(x,y,z,w)$.

Note that, when the two centers of merons coincide as $l=0$, 
this two-merons system becomes an instanton with the size $a$, 
which corresponds to a single baryon in holographic QCD.
In fact, this two-merons configuration
is a special deformation of an instanton, and 
the two-merons oscillation in holographic QCD
can be regarded as a special excitation 
of a single baryon.
Then, the two-merons system has a finite energy like an instanton in holographic QCD.
Here, we take the meron size to be $a \simeq 0.4{\rm fm}$, 
corresponding to the instanton size  
describing the ground-state baryon in holographic QCD 
up to the next leading of $1/\lambda$, 
as was shown in the previous section.

To get an analytical rough estimate for the two-merons oscillation energy, 
we consider $S_{\rm 5YM}$, the leading term on $1/N_c$ and $1/\lambda$, 
and use an approximation of $h(w)=k(w)=1$.
After a lengthy calculation, we obtain the Lagrangian for $l(t)$ up to $O(l^4)$:
\begin{eqnarray}
L[l(t)]=\frac{1}{2} m l^2 \dot l^2-\frac{1}{2} k l^2-\frac{1}{8} m \omega^2 l^4,
\end{eqnarray}
with the coefficients of 
\begin{eqnarray}
m=\frac{216}{5} \pi^2 \kappa a^{-2}, \ \
k=\frac{12}{5} \pi^2 \kappa a^{-2}, \ \
\omega =\frac{1}{6} \sqrt{\frac{202}{7}} a^{-1} \simeq 0.9 a^{-1}.
\end{eqnarray}
By changing the variable $q\equiv \frac{1}{2}l^2$, 
we find 
\begin{eqnarray}
L=\frac{1}{2} m \dot q^2-k q-\frac{1}{2} m \omega^2 q^2.
\end{eqnarray}

Thus, the oscillation energy of this meron-pair is finite, 
and its rough estimate at the semi-classical level is 
about $\omega \simeq 450{\rm MeV}$ for $a \simeq 0.4{\rm fm}$ 
for the lowest excitation mode, 
although there is some uncertainty on the boundary condition at $q=0$. 

Because of the oscillation in the extra $w$-direction,
all the quantum numbers of this-type excited baryons 
are just the same as the ground-state baryon. 
Then, as a possibility, this excitation might correspond to 
the Roper resonance ${\rm N}^*(1440)$.

Note that this oscillation is in the extra $w$-direction, 
which is independent of ordinary coordinates $(x,y,z)$.
Then, as an interesting conjecture, 
this type of oscillating excitation in the extra $w$-direction 
might universally appear for any baryon, 
keeping the same quantum numbers. 

Such an oscillation mode in the extra $w$-direction 
is easily grasped in the holographic description, although 
it would be described as a highly complicated collective mode in QCD 
without viewpoint of the extra direction.

\section{Summary and Concluding Remarks}

In this paper, we have studied topological objects in holographic QCD in the two-flavor ($N_f=2$) case based on the Sakai-Sugimoto model, 
which is infrared equivalent to 1+3 dimensional massless QCD. 
Using the gauge/gravity duality, holographic QCD is described as 
1+4 dimensional U($N_f$) gauge theory in flavor space with a background gravity, 
and its instanton solutions correspond to baryons. 

First, we have reduced 1+4 dimensional holographic QCD into a 1+2 dimensional Abelian Higgs theory in a curved space using the Witten Ansatz, 
and have considered its topological aspect. We have numerically calculated 
the Abrikosov vortex solution and have investigated single baryon properties. 

Second, we have studied a single meron and two merons in holographic QCD. We have found that the single meron carrying a half baryon number has an infinite energy also in holographic QCD. We have proposed a new-type baryon excitation of the two-merons oscillation in the extra-direction of holographic QCD. 

\begin{figure}[h]
\centerline{\includegraphics[width=12cm]{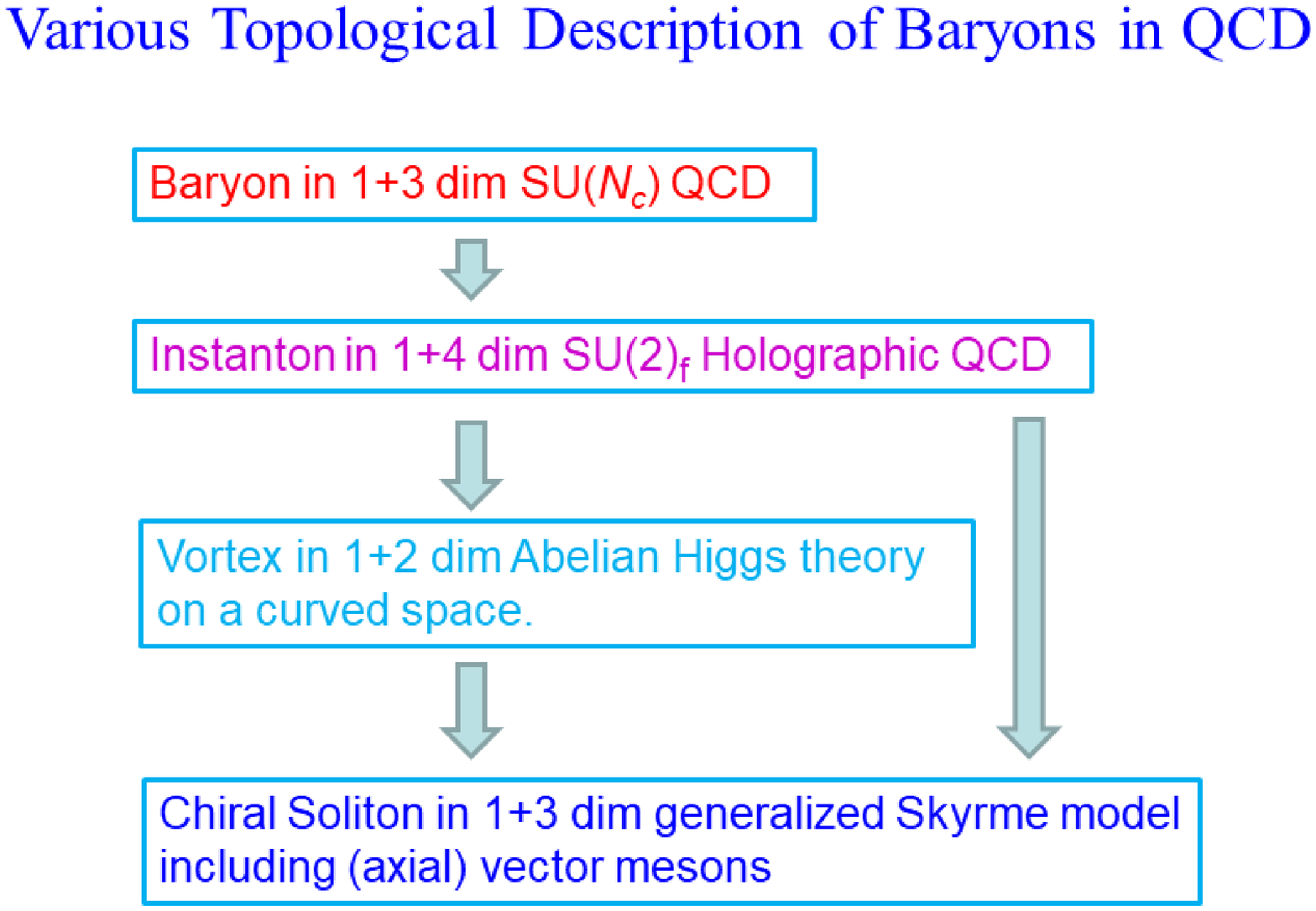}}
\caption{Various topological description of baryons in QCD. In 1+4 
holographic QCD (HQCD), baryons are described with instantons 
in four-dimensional space $(x,y,z,w)$.
The baryon is also expressed as the Abrikosov vortex 
in the 1+2 Abelian Higgs theory reduced from HQCD.
The baryon is eventually described as a topological chiral soliton in 
a 1+3 generalized Skyrme model derived from HQCD.
These different topological objects are mathematically related, 
and these topological charges are exactly identical.
}
\label{fig:sum}
\end{figure}

In the end, we summarize in Fig.\ref{fig:sum} 
various topological description of baryons presented from holographic QCD: 
\begin{itemize}
\item
In 1+4 dimensional SU($N_f$) holographic QCD, 
baryons are described with instantons in four-dimensional space $(x,y,z,w)$, 
based on the nontrivial homotopy group $\Pi_3({\rm SU}(N_f))={\bf Z}$.
\item
In the 1+2 dimensional Abelian Higgs theory reduced from holographic QCD, 
the baryon is expressed as the Abrikosov vortex in $(r,w)$-plane,
based on $\Pi_1({\rm U}(1))={\bf Z}$.
\item
The baryon is also described as a topological chiral soliton 
in a 1+3 dimensional generalized Skyrme model derived from holographic QCD, 
based on 
$\Pi_3({\rm SU}(N_f)_L \times {\rm SU}(N_f)_R/{\rm SU}(N_f)_V)={\bf Z}$.
\end{itemize}
These different topological objects are mathematically related, 
and these topological charges are exactly identical. 
In this way, holographic QCD gives a new physical picture and deep insight 
for baryons in QCD.

\section*{Acknowledgements}
We thank Prof. S. Sugimoto and Dr. T. Ishii for their useful comments and discussions.
H.S. is supported by the Grants-in-Aid for
Scientific Research [19K03869] from Japan Society for the Promotion of Science.

\vspace{15pt}
\noindent
{\bf References}
\vspace{15pt}


\end{document}